\documentclass[pdflatex,sn-nature,iicol]{sn-jnl}

\usepackage{eso-pic}

\makeatletter
\def\Authorfont{\reset@font\fontsize{10bp}{12bp}\selectfont\boldmath\titraggedcenter}
\def\addressfont{\reset@font\fontsize{9bp}{11bp}\selectfont\titraggedcenter}
\makeatother

\usepackage{graphicx}
\usepackage[dvipsnames]{xcolor}
\usepackage{amsmath,amssymb}
\usepackage{booktabs}
\usepackage{braket}

\usepackage[T1]{fontenc}
\usepackage[normalem]{ulem}
\begin{document}

\title {Learning non-Markovian noise on a quantum processor}

\author[1]{\fnm{Jasleen} \sur{Kaur}}

\author[2]{\fnm{Abhinash Kumar} \sur{Roy}}

\author[1]{\fnm{Alexei} \sur{Gilchrist}}
    
\author[1]{\fnm{Christina} \sur{Giarmatzi}}

\affil[1]{\orgdiv{School of Mathematical and Physical Sciences},
\orgname{Macquarie University},
\orgaddress{\city{Sydney}, \state{NSW}, \country{Australia}}}

\affil[2]{\orgdiv{Science, Mathematics and Technology Cluster},
\orgname{Singapore University of Technology and Design},
\orgaddress{\city{8 Somapah Road, 487372}, \country{Singapore}}}

\abstract{ 
Complex time-correlated (non-Markovian) noise can significantly influence the dynamics of quantum devices. However, characterising this class of noise typically requires full multi-time process tomography, which is experimentally and computationally expensive,  therefore difficult to use as a routine diagnostic tool. In this work, we develop a resource efficient, neural network based approach, for estimating the amount of non-Markovianity. We demonstrate it on a superconducting quantum processor, and show that observations obtained using a single causal-break instrument support accurate 
quantitative prediction of non-Markovianity. Furthermore, we show that the non-Markovianity range obtained through maximum-likelihood reconstruction from restricted statistics is less discerning in comparison. Our results provide a practical route towards repeated diagnostic assessment of temporal correlations with reduced experimental requirements.
}
\AddToShipoutPictureFG*{%
    \AtTextLowerLeft{%
        \raisebox{-6mm}{%
            \footnotesize
            \href{mailto:jasleen.kaur8@hdr.mq.edu.au}
            {jasleen.kaur8@hdr.mq.edu.au}
        }
    }
}
\maketitle

Noise limits the reliability and scalability of today’s quantum devices and is a major barrier to fault-tolerant quantum computing \cite{Preskill2018,Campbell2017}.
Conventional noise models used in device characterization and error correction often assume 
independent errors or memoryless (Markovian) noise 
\cite{Knill1998, BlumeKohout2025QCVV,Burgelman2025,Terhal2015}. 
However, non-Markovian noise, where past system-environment interactions affect future behavior, has been observed in current quantum hardware \cite{Wei2024,McEwen2021,White2020,Giarmatzi2025,Oda2026,Arute2019,Eickbusch2026}. 
This noise can lead to temporally correlated errors that affect the design of control \cite{White2020,White2022,White2025}, mitigation, and correction strategies 
\cite{Kam2025,LiuXiaoCai2024,Srivastava2026}. 
In this sense, characterizing non-Markovian noise is therefore an important step towards complete characterisation of quantum devices, complementing standard reported metrics
such as coherence times and gate and readout errors, which do not capture or quantify non-Markovian noise \cite{IBMCalibration, GoogleWillow2024,Wei2024}.

A variety of approaches have been developed to characterize non-Markovian dynamics \cite{Breuer2016,LI20181}. A widely used approach employs dynamical maps, which encode general transformations and use several criteria, such as information backflow and CP-divisibility \cite{Breuer2009,Rivas2010}, to determine whether the environment acts as a memory leading to non-Markovian effects. 
However, this approach captures only two-time correlations, is limited in its inference of memory, for instance, CP-divisibility does not imply Markovianity \cite{Milz2019}, 
and fails in scenarios with initial system-environment correlations \cite{Pechukas1994,Modi2012}.

The process matrix framework developed in \cite{Oreshkov2012}, or process tensors \cite{Pollock2018a,Pollock2018b},
overcomes the aforementioned limitations. 
In this framework, a quantum system undergoing  transformations subject to non-Markovian noise is a \textit{multi-time quantum process} described by a process matrix. Similar to a density matrix that fully describes a quantum state, a process matrix fully describes a multi-time quantum process \cite{Shrapnel_2018}, including its non-Markovian noise. Similarly to reconstructing a density matrix from measurement data, reconstructing a process matrix (and its non-Markovianity) requires to perform operations at each timestep of the multi-time process.
In \cite{Giarmatzi2025}, full multi-time process tomography has been used to demonstrate the presence of non-Markovian noise on 
superconducting processors.
However, full tomography to reconstruct the process matrix requires performing an informationally complete set of operations at each timestep, making it a resource-intensive task \cite{White2022, roy2026practical}.

Alternatively, a technique using only sequence of unitaries as the operations at the intermediate timesteps followed by a final measurement has been used to investigate non-Markovian noise \cite{Milz2018LimitedControl,White2020,White2022}. However, it provides a restricted characterization of the multi-time process, as it constrains the amount of non-Markovianity, defined as the distance to the closest Markovian process, only through upper and lower bounds, rather than providing a direct quantitative estimate \cite{White2025Unitary}. 
Therefore, the development of resource efficient alternatives to infer non-Markovianity while using fewer experimental and computational resources is highly desirable.

Recently, machine learning algorithms have been explored as resource efficient tools for characterizing and quantifying temporally correlated noise \cite{Fanchini2021,Scarpetta2025}. 
For instance, a supervised learning algorithm has been used to estimate the dimension of the environment in numerically simulated non-Markovian processes \cite{Shrapnel2018}. This required less data than
full tomography, which we call an \textit{informationally incomplete} setting.
Also, in \cite{Goswami2021}, it was demonstrated that AI can learn the amount of non-Markovianity without full tomography in a quantum optical experiment.
However, the demonstration involved a restricted class of processes, known as Classical Common Cause (CCC) processes \cite{giarmatzi2021witnessing}, where they inject correlated classical noise on the intermediate steps of a multi-time process. This leaves open the question of whether such an approach can generalize to a broader and more realistic class of non-Markovian noise which is naturally present in a quantum processor.

In this work, we develop a machine learning approach using neural networks to quantify naturally occurring non-Markovianity in quantum processors using informationally incomplete experimental data. We demonstrate 
that observations obtained using a single operation at the intermediate timestep enable accurate prediction of non-Markovianity, benchmarked against values obtained from multi-time process tomography. 
We compare our results with the restricted process characterization using maximum likelihood
and show that the restricted data admits substantially broader range of non-Markovianity, while our neural network closely predicts the true value obtained from tomography.
These results demonstrate a scalable method to accurately estimate non-Markovian noise in a multi-time process, providing a practical route to incorporating non-Markovianity into routine characterization of current quantum devices.

\section{Results}

\textbf{Multi-time process formalism:} The process matrix 
formalism provides
a general framework to  
describe correlations between parties
with no assumption of a pre-existing background time or global causal structure \cite{Oreshkov2012}. When a definite causal order is assumed, as in the case of ordinary quantum circuits, the process matrix reduces to  quantum combs \cite{Chiribella2009} and directly relates to the process tensor framework \cite{Pollock2018a,Pollock2018b}. The framework provides a way to fully characterize a multi-time process, 
through a series of operations at each timestep of the process. In this way, it provides an operational way to characterize non-Markovianity of a multi-time process
\cite{Costa2016, Pollock2018a}.

\begin{figure}[h!]
    \centering
\includegraphics[width=\columnwidth]{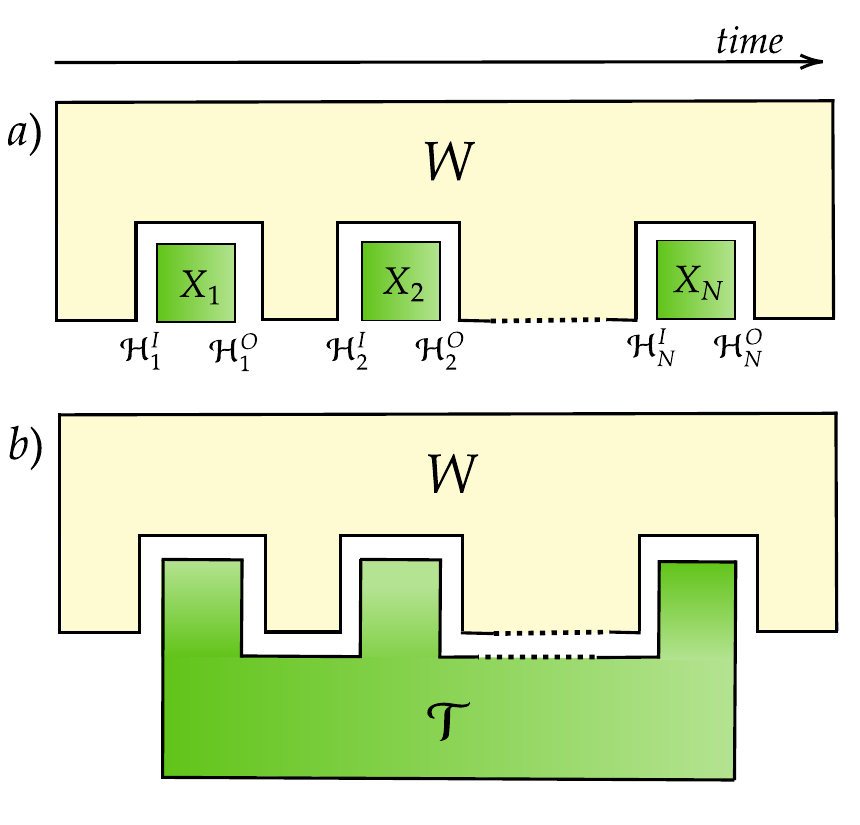}
    \caption{\textbf{Schematic representation of a multi-time process with $N$ time steps} (a) $X_1,X_2,\ldots,X_N$ denote the time steps at which the experimenter can perform a quantum operation on the system. Each timestep is associated with input and output Hilbert spaces $\mathcal{H}_i^I$ and $\mathcal{H}_i^O$, respectively. Time progresses from left to right. (b) The operations implemented by the experimenter at different time steps to probe the process are collectively represented by the tester $\mathcal{T}$, which could in general represent correlated operations.
    The process matrix $W$ describes all aspects of the underlying dynamics, except $\mathcal{T}$. }
    \label{fig:process_matrix}
\end{figure}

To probe the dynamics of the system evolving in time, we perform interventions at a sequence of times, which we call \emph{labs} (see Fig.~\ref{fig:process_matrix}). 
Each such lab $X_i$ is associated with input and output Hilbert spaces 
$\mathcal{H}_i^{I}$ and $\mathcal{H}_i^{O}$, respectively, in which arbitrary operations may be performed on the system. 
These operations are represented by completely positive, trace non-increasing maps $\mathcal{M}_{a|x}:\mathcal{L}(\mathcal{H}_i^I)\rightarrow\mathcal{L}(\mathcal{H}_i^O)$, where $x$ is 
the setting, $a$ is 
the outcome,  and $\mathcal{L}(\mathcal{H})$ denotes the space of bounded linear operators on the Hilbert space $\mathcal{H}$. For each setting $x$, the collection $\{\mathcal{M}_{a|x}\}_a$ forms a \emph{quantum instrument} provided that $\sum_a\mathcal{M}_{a|x}$ is trace preserving. In the following, we use the Choi representation of maps, where the Choi operator of $\mathcal{M}_{a|x}$ is defined as
\begin{equation}
M_{a|x}
=
\sum_{j,k=1}^{d_i^I}
|j\rangle\langle k|
\otimes
\mathcal{M}_{a|x}\!\left(|j\rangle\langle k|\right),
\end{equation}
with $\{|j\rangle\}_{j=1}^{d_i^I}$ a fixed orthonormal basis of the input space $\mathcal{H}_i^I$. 

Consider a sequence of operations, with $M^{X_i^I X_i^O}_{a_i|x_i}$ denoting the Choi operator of the operation associated with a setting $x_i$ and an outcome $a_i$ at lab $X_i$. Defining $\mathbf{a}=(a_1,\ldots,a_N)$ and $\mathbf{x}=(x_1,\ldots,x_N)$, the corresponding outcome statistics are determined by the generalized Born rule \cite{Shrapnel2018BornRule},
\begin{equation}
p(\mathbf{a}|\mathbf{x})
=
\operatorname{Tr}\!\left[
W^T
\bigotimes_{i=1}^{N}
M^{X_i^I X_i^O}_{a_i|x_i}
\right]=
\operatorname{Tr}\!\left(
W^T
\mathcal{T}_{\mathbf{a}|\mathbf{x}}
\right),
\label{eq:generalized_born_rule}
\end{equation}
where superscript $T$ denotes the transpose operation, the operators $W$ and $\mathcal{T}_{\mathbf{a}|\mathbf{x}}$  are known as the process matrix and the tester, respectively.

The process matrix describes the underlying process, including any initial system-environment correlations and the system-environment dynamics between the labs. It is a positive semidefinite operator on the joint input and output Hilbert spaces of the labs, i.e., 
\begin{equation}
W\in
\mathcal{L}\!\left(
\bigotimes_{i=1}^{N}
\mathcal{H}_{i}^{I}\otimes\mathcal{H}_{i}^{O}
\right).
\label{eq:W_IO_space}
\end{equation}
In addition, $W$ is constrained by causality conditions that ensure consistency with the causal order of operations \cite{Araujo_2015,Costa2016}.

In this work, the \emph{tester} is given by the tensor product of Choi operations in the labs. It represents the collective operation in the labs associated with setting $\mathbf{x}$ and outcome $\mathbf{a}$. Different tester sets therefore correspond to different sets of operations used to probe the same underlying process. \\ \\
\textbf{Quantifying non-Markovianity:} In Markovian processes, the time evolution of a quantum system is memoryless. The statistics observed at each timestep  depend only on the previous one.
By contrast, non-Markovian dynamics induce  memory effects. During the evolution of a system between timesteps, there can be system-environment interactions. 
These interactions can be correlated in time because the environment can act as a carrier of memory, taking the information about each interaction forward in time. The existence of these correlations is encoded in the structure of the process matrix. In particular, a  Markovian process matrix factorizes into a tensor product of channels associated with successive time steps. 
For a multi-time process with $N$ time steps, this takes the form
\begin{equation}
W_{Markov}
=
\rho^{X_1^I}
\otimes
\left(
\bigotimes_{i=1}^{N-1}
T^{X_i^O X_{i+1}^I}
\right)
\otimes
\mathbb{I}^{X_N^O},
\end{equation}
where $\rho^{X_1^I}$ 
denotes the state of the system associated with the input space of the first timestep, $T^{X_i^O X_{i+1}^I}$ denotes the Choi operator of the channel connecting two consecutive timesteps  $X_i$ and $X_{i+1}$, and $\mathbb{I}^{X_N^O}$ denotes the identity operator on the output space of the final timestep. If such a factorization is not possible, the process is  non-Markovian. Hence, the process matrix formalism provides a natural way to infer the presence of non-Markovianity.

In this work, non-Markovianity is quantified using the relative entropy distance of the process matrix \(W\) and its  closest Markovian  \(W_M\). 
\begin{equation}
\mathcal{N}(W)
=
\min_{W_M \in \mathcal{W}_M}
S(w\|w_M),
\label{eq:RE_measure}
\end{equation}
where \(\mathcal{W}_M\) 
denotes the set of Markovian process matrices, and the lowercase $w = W/\operatorname{Tr}(W)$ and $w_M = W_M/\operatorname{Tr}(W_M)$ are processes normalized to one for the relative entropy to be well-defined. 

The quantum relative entropy between two states $w_1$ and $w_2$ is given by
\begin{equation}
   S(w_1\|w_2)
=
\mathrm{Tr}
\left[
w_1
\left(
\log w_1
-
\log w_2
\right)
\right]. 
\end{equation}
Importantly, the above optimization admits a closed form solution, with the minimum attained by the Markovian product process constructed from marginals of $W$ (see Methods for details). However, as evident, the evaluation of non-Markovianity through this measure requires knowledge of the process matrix $W$, which can be reconstructed through multi-time process tomography.\\ \\
\textbf{Multi-time process tomography: } 
The process matrix has a multipartite structure across the input and output spaces of the labs, allowing ideas from state tomography to be extended to multi-time process tomography. 
Just as the density matrix of a state can be reconstructed by the probabilities of an Informationally Complete (IC) positive operator-valued measure, analogously, the process matrix of a multi-time process can be reconstructed by the probabilities obtained from operations performed at multiple timesteps, see Eq.~\eqref{eq:generalized_born_rule}. If the operations performed at each  timestep form an IC basis, then the combined multi-time data can determine the process matrix $W$ uniquely.  
For the operations to form an IC basis, it means that the operations span the relevant operator space associated with the process matrix. Conversely, if the operations span only a \emph{restricted} subspace, 
the resulting data is \emph{informationally incomplete} and does not uniquely determine the full process matrix. An example of a restricted set is when only unitary operations are performed at a timestep, which span a $(d^2-1)^2+1$ dimensional space, instead of a $d^4$ dimensional space required for the Choi matrices of the operations on a $d$ dimensional system \cite{Milz2018LimitedControl, roy2026practical}. 
This means that the experimental data obtained with such restricted operations are insufficient to uniquely reconstruct $W$. 

In this work, such informationally incomplete experimental datasets, consisting of probabilities obtained from restricted sets of operations performed at different time steps, serve as \textit{features} in our machine learning analysis. 
We therefore ask whether such restricted experimental information retains sufficient information to infer the amount of non-Markovianity present in the multi-time process we investigate.

\subsection{Experiment}

\begin{figure}[h!]
    \centering
\includegraphics[width=\columnwidth]{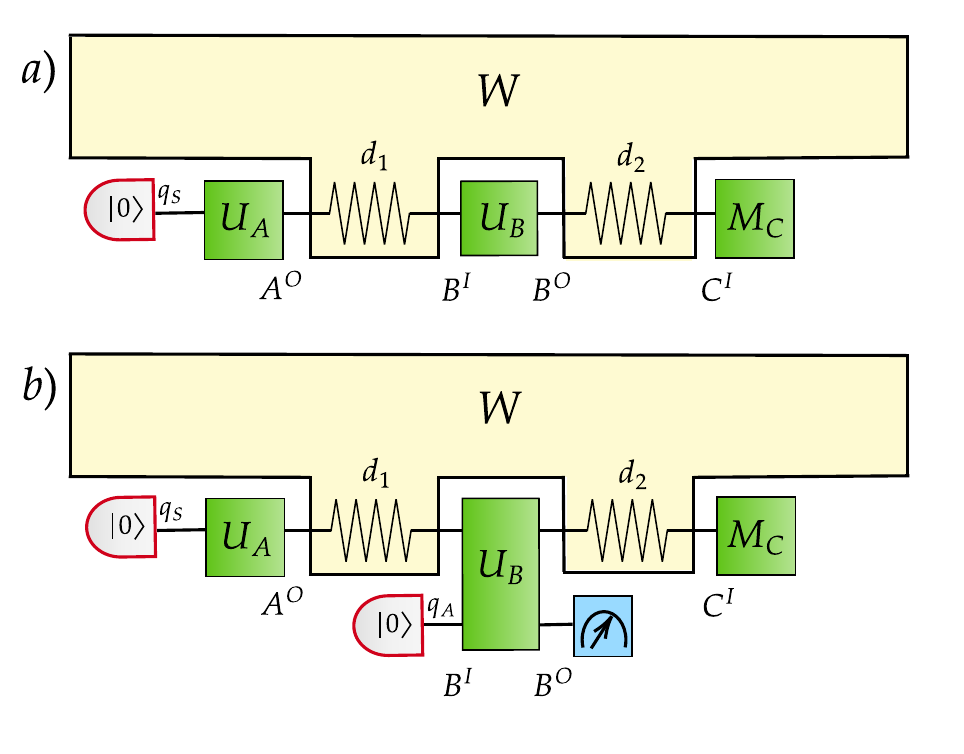}
    \caption{\textbf{Representative circuits for the multi-time process characterization.} The system qubit $q_S$ is probed sequentially at labs $A$, $B$, and $C$, corresponding to state preparation, intermediate intervention, and final measurement, respectively. These operations are separated by controlled idle intervals $d_1$ and $d_2$. The process is represented as $A^O \rightarrow B^IB^O \rightarrow C^I$, where the superscripts $I$ and $O$ denote the input and output spaces of a lab, respectively. (a) A circuit involving only the system qubit, used for deterministic operations at the intermediate time step. (b) An ancilla-assisted circuit used to implement non-deterministic operations at $B$.
    }
    \label{fig:experimental_protocol}
\end{figure}

To probe non-Markovianity on the processor, we experimentally characterize multi-time quantum processes on different qubits.  
Each qubit undergoes three successive operations at the labs \(A\), \(B\), and \(C\). These are separated by two controlled idle intervals $d_1$ and $d_2$. By varying these intervals, we realize multi-time processes corresponding to different evolution times.
A schematic representation of the experimental protocol is shown in Fig.~\ref{fig:experimental_protocol}.

For each choice of system qubit and delay intervals $(d_1,d_2)$, an informationally complete set of operations is implemented at each of the three timesteps to perform multi-time tomography. 
obtained probabilities are used to reconstruct 
the process matrix $W$ using maximum likelihood estimation (MLE). The reconstructed process is then compared with its closest Markovian process using the relative-entropy measure given in Eq.~\ref{eq:RE_measure},
 yielding a non-Markovianity value $\mathcal{N}$ for each experimentally realized process.

The experiments are performed on the 156-qubit IBM superconducting processor \texttt{ibm\_fez}.
To optimize hardware access time, we perform multiple tomography experiments in parallel across the processor (see Methods).
The complete dataset contains 2304 experimentally characterized processes obtained across different delay intervals $(d_1,d_2)$ and physical qubits, therefore, different physical locations on the processor. The resulting $\mathcal{N}$ values provide the 
\emph{labels} of non-Markovianity used throughout the subsequent machine learning analysis.

\subsection{Machine learning in informationally incomplete regime}
For the set of multi-time processes under consideration, 
our choice of complete tomographic set comprises four linearly independent preparations in lab $A$, sixteen linearly independent operations in lab $B$, and measurement in three Pauli basis in lab $C$ resulting in six operations, making this choice overcomplete. 
Informational completeness at the intermediate timestep $B$ requires nondeterministic operations which are implemented using an ancillary qubit. The complete operation sets, along with their circuit and chip level implementations, are described in Methods. 
Overall, reconstruction in informationally complete regime requires $156$ tomographic circuits 
and yields $228$ independent probabilities for each process. 

We contrast this with two informationally incomplete tester sets, each retaining a single \textit{causal-break} 
instrument at $B$, which measures the incoming state and prepares a fixed output state. The first informationally incomplete tester set consists of four linearly independent preparations at $A$ and three measurement settings at $C$, yielding $36$ independent probabilities, while the second consists of a single preparation at $A$ and a single measurement at $C$, 
yielding only $3$ 
independent probabilities. These 
two tester sets 
require only $12$ and $1$ tomographic circuits to be implemented, respectively, compared to $156$ for the full IC reconstruction. 

To learn a direct mapping from the experimentally measured probabilities to the IC-derived $\mathcal{N}$ values without reconstructing the full process matrix, we employ a feed-forward artificial neural network (ANN).
We first evaluate the estimator in the \textit{interpolation} regime, where 
$2304$ experimentally characterized processes are randomly partitioned into training, validation, and test sets. The $36$-feature tester achieves a mean absolute error of \(0.01352\pm0.00036\) and \(R^2=0.9701\pm0.0054\), while the highly restricted $3$-feature tester achieves an MAE of \(0.02073\pm0.00082\) and \(R^2=0.9092\pm0.0246\) (Table~\ref{tab:iic_performance}). 

\begin{table*}[h!]
\centering
\caption{\textbf{Performance of the ANN for two informationally incomplete tester sets.}
The MAE and $R^2$ values are reported as a mean $\pm$ a standard deviation, over four random seeds.
For interpolation, both the data partition and ANN training choices are varied,
whereas for forecasting, the delay-defined partition is fixed and the ANN training seed is varied.}
\label{tab:iic_performance}

\resizebox{0.97\textwidth}{!}{
\begin{tabular}{c ccc c cc cc}
\toprule
& \multicolumn{3}{c}{No. of Instruments}
&
& \multicolumn{2}{c}{Interpolation}
& \multicolumn{2}{c}{Forecasting} \\

\cmidrule(lr){2-4}
\cmidrule(lr){6-7}
\cmidrule(lr){8-9}

\shortstack{Tester\\sets}
& Lab A
& Lab B
& Lab C
& Circuits
& MAE
& $R^2$
& MAE
& $R^2$ \\

\midrule

36-feature
& 4
& 1
& 3
& 12
& $0.01352 \pm 0.00036$
& $0.9701 \pm 0.0054$
& $0.01747 \pm 0.00278$
& $0.9674 \pm 0.0116$ \\

3-feature
& 1
& 1
& 1
& 1
& $0.02073 \pm 0.00082$
& $0.9092 \pm 0.0246$
& $0.02448 \pm 0.00058$
& $0.9367 \pm 0.0086$ \\

\bottomrule
\end{tabular}
}

\end{table*}

\begin{figure*}[h!]
    \centering
    \includegraphics[width=0.63\textwidth]{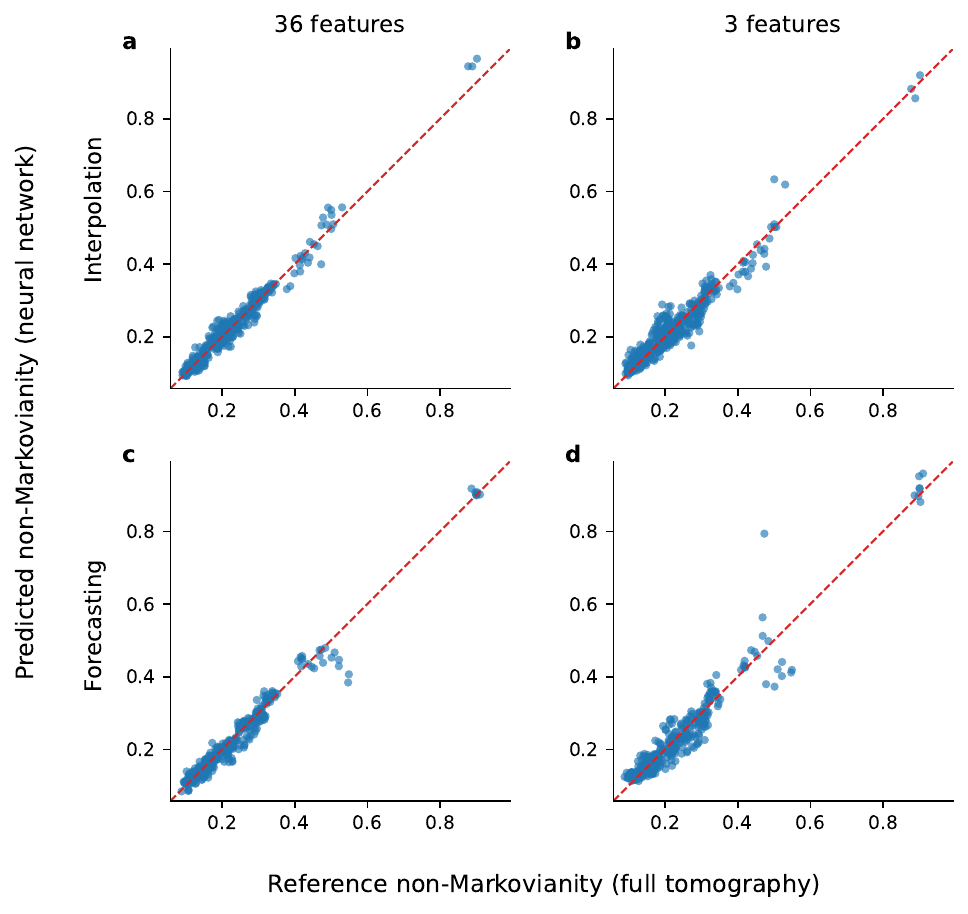}
    \caption{\textbf{ANN estimation of non-Markovianity under interpolation and forecasting.} ANN estimates of non-Markovianity are compared with the corresponding reference values obtained from informationally complete reconstruction. For the \(36\)-feature tester with, (a) interpolation, and (c) forecasting; and for the \(3\)-feature tester with, (b) interpolation, and (d) forecasting. The dashed line denotes perfect agreement between the ANN estimate and the full tomography reference value.} 
    \label{fig:iic_36_3}
\end{figure*}

These results show that, although the restricted testers are informationally incomplete and cannot uniquely reconstruct the full process matrix, they retain sufficient information to reliably predict the amount of non-Markovianity. Notably, substantial predictive performance is retained even when the input is reduced to only three independent probabilities. In other words, we show that probabilities obtained using a single causal-break 
operation at the intermediate timestep is sufficient to estimate non-Markovianity with high predictive accuracy.

We next test whether the learned mapping generalizes beyond the delay regime represented in the training data. For this temporal \textit{forecasting} task, the ANN is trained on processes with shorter total delay durations and evaluated on processes with longer, previously unseen total delays. It should be noted here that the delay values are used exclusively to define the split and are never supplied to the ANN as input features. Despite this more stringent separation, the $36$-feature tester achieves an MAE of $0.01747\pm0.00278$ and $R^2=0.9674\pm0.0116$, while the $3$-feature tester achieves an MAE of $0.02448\pm0.00058$ and $R^2=0.9367\pm0.0086$ (Table~\ref{tab:iic_performance}). Thus, the estimator retains strong predictive performance when applied to processes at longer total delays that were not represented during training, indicating that the learned relationship between the restricted tester probabilities and $\mathcal{N}$ extends beyond the temporal regime used to train the model.

Figure~\ref{fig:iic_36_3} shows the ANN-predicted values of $\mathcal{N}$ against the corresponding reference values, obtained from tomography, for both tester configurations under the interpolation  and forecasting regimes.
We remind that, in the interpolation regime, the training set included the processes with the full range of total durations of the two idle intervals. In contrast, in the forecasting regime, the training set included the processes with shorter total durations, while predicting the non-Markovianity for processes with longer, previously unseen total durations. 
In all cases, the predictions remain closely aligned with the diagonal 
(perfect agreement of the ANN with the tomography results), with the $36$-feature tester exhibiting the tightest agreement. The $3$-feature tester shows a larger spread, as expected from the 
extremely reduced information available to the model, but still preserves a clear correspondence with the reference values in both evaluation settings. 

Together with the performance metrics in Table~\ref{tab:iic_performance}, these results demonstrate that a very small subset of
tomographic data can still 
provide a reasonable estimate of 
non-Markovianity 
using a neural network, and the learned relation survives a temporal-distribution shift 
of the idle times of the processes. This provides a practical route towards incorporating non-Markovianity into repeated hardware-characterization workflows where experimental overhead is a limiting factor.  

\subsection{Comparative analysis: Restricted MLE vs AI}
In the preceding section, we showed that neural networks can accurately predict non-Markovianity using only a small subset of the informationally complete measurement data. To place this predictive accuracy in context, we now examine what can be inferred from the same restricted observations within the maximum-likelihood reconstruction framework \cite{White2022}. Since these observations admit multiple compatible physical processes, 
we numerically explore the range of non-Markovianity across this feasible set, and compare the resulting range with the neural-network predictions and the reference values obtained from informationally complete tomography.

To this end, we use the 36-feature tester introduced above, comprising four preparations at laboratory $A$, a single causal-break instrument at $B$, and three measurement settings at $C$, with the intermediate and final outcomes retained in the joint statistics. We focus on this configuration because the three-feature tester uses a subset of these settings, and reducing the features can only enlarge the set of compatible physical processes. Consequently, any ambiguity in non-Markovianity established for the 36-feature tester also persists for the more restricted three-feature tester.

Consider the joint tester operators, $\{\mathcal{T}_r\}_{r\in R}$, associated with the restricted configuration, where $r$ labels the retained settings and outcomes. The restricted fitted probabilities $\{\hat{p}_r\}_{r\in R}$ define the following set of compatible process matrices
\begin{equation}
\mathcal{C}_R
=
\left\{
W\in\mathcal{W}:
\operatorname{Tr}(W\mathcal{T}_r)=\widehat{p}_r
\quad \forall r\in R
\right\},
\end{equation}
where $\mathcal{W}$ is the set of process matrices satisfying positivity, normalization, and the appropriate causality constraints. The fitted probabilities $\widehat{p}_r$ are obtained from $\widehat{p}_r
=
\operatorname{Tr}\!\left(
{W}_{\mathrm{IC}}\mathcal{T}_r
\right),$
with ${W}_{\mathrm{IC}}$ denoting the maximum-likelihood reconstruction from the informationally complete data. Thus, $\mathcal{C}_R$ contains all physical processes that reproduce the restricted statistics predicted by this reconstruction, including $\widehat{W}_{\mathrm{IC}}$ itself. It is important to note that here we use fitted probabilities rather than experimental frequencies because finite-shot fluctuations can make the latter incompatible with exact physical and causality constraints. This construction ensures a nonempty compatible set and allows us to examine the ambiguity 
in non-Markovianity arising from the restricted measurement configuration.
\begin{figure}[!h]
    \centering
    \includegraphics[width=0.45\textwidth]
    {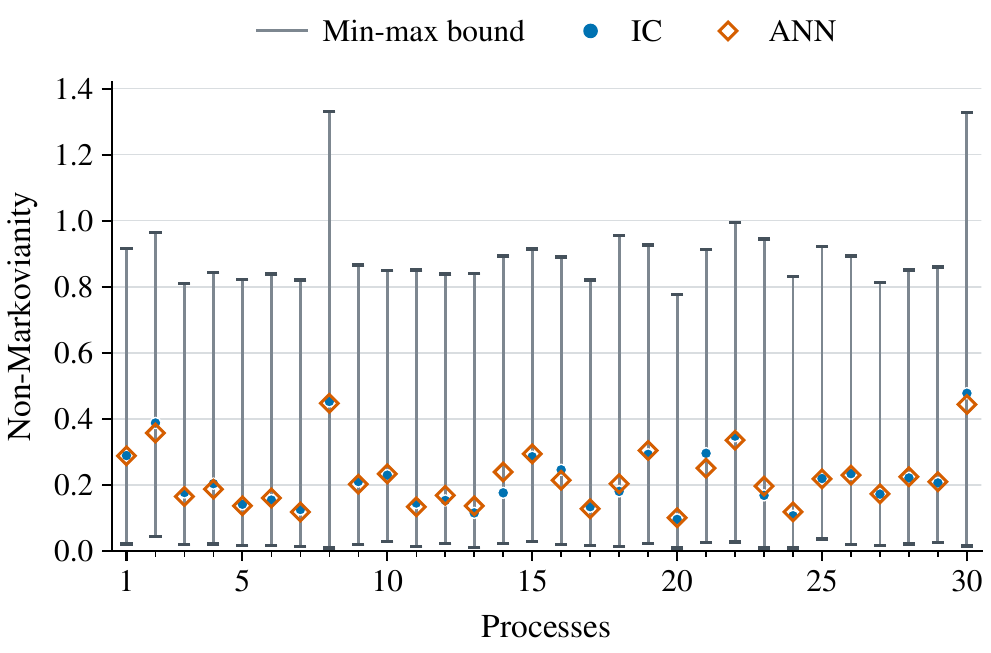}
    \caption{\textbf{Relative-entropy non-Markovianity ($\mathcal{N}$) for 30 randomly selected test processes.} Blue circles show references obtained from informationally complete MLE reconstruction; orange open diamonds show neural network predictions using the 36-feature tester. Grey intervals span the lowest and highest non-Markovianity values attained in a numerical search over the feasible set of processes consistent with the restricted feature probabilities. The broad compatible ranges show that the restricted data alone do not tightly constrain non-Markovianity, while ANN estimates remain close to the IC reference values. }
    \label{fig:RE_comparison}
\end{figure}

We explore the variation of relative entropy 
as a measure of 
non-Markovianity within $\mathcal{C}_R$ using an entropy-based iterative optimization (See Methods for details). 
In Fig. \ref{fig:RE_comparison}, we compare these ranges with the informationally complete MLE reference values and neural network predictions for selected test processes. We observe that the restricted statistics admit a large range of values not centered around the reference ones, yet the neural network predictions remain close to the references. 
These results show that the ANN can learn patterns in the experimental data to accurately predict non-Markovianity, even when the restricted observations do not uniquely determine it.

\section{Conclusion}
We demonstrate a neural-network approach for estimating non-Markovianity 
of multi-time processes on a superconducting quantum processor using informationally incomplete 
tomographic data.
The multi-time processes are characterized by operations in three successive labs, involving two idle intervals in between the labs.
Remarkably, even the probabilities obtained from a single circuit 
involving a causal-break instrument in the intermediate lab enable accurate predictions of non-Markovianity under both interpolation and forecasting regimes.
In the interpolation regime, the processes with the full range of total durations of the two idle intervals are used for training. However, in the forecasting regime, the processes with shorter total durations are used for training, while predicting the non-Markovianity for processes with longer total durations.

Moreover, for the same informationally incomplete data, 
MLE reconstruction reveals 
a large range of non-Markovianity. 
Nevertheless, the neural network predictions 
closely agree with the values obtained from full multi-time process tomography. 

Our approach is particularly relevant to routine device characterization techniques. Following an initial investment in tomographically obtained training labels, restricted measurements and neural-network inference 
reduce the resources required for subsequent assessments. Importantly, reasonable predictive accuracy is achieved with as few as 400 experimentally characterized processes in the training set. 
This provides a route towards incorporating non-Markovianity into routine characterization alongside coherence times and gate and readout errors.

We note that, in our tests, training on randomly generated simulated processes did not yield accurate predictions for the experimental processes, which highlights the importance of the training 
set. 
However, a suitable 
noise model could enable training on simulated data, substantially reducing the need for experimental training labels.
Given the efficacy of our approach, it would be interesting to explore whether observations obtained using a single causal-break instrument retain their predictive power for processes involving more timesteps.  

\section{Methods}
\subsection{Circuit level implementation}\label{Methods::circuit & chip}

The multi-time process under consideration is $A^O \rightarrow B^IB^O \rightarrow C^I$. (Fig.~\ref{fig:experimental_protocol}). At laboratory \(A\), an informationally complete set requires $4$ linearly independent operations. Hence, a set of input states
$
\{\ket{0},\ket{1},\ket{+},\ket{i}\}
$
is prepared using appropriate single-qubit rotations. Following state preparation, the system undergoes a period of free evolution implemented via a controlled delay for $d_1$ units, 
a unit used by IBM denoting $4 \mathrm{ns}$. 

At the intermediate laboratory \(B\), an informationally complete set consisting of $16$ linearly independent operations is implemented. In particular, ten deterministic single-qubit unitary operations are used:
$ \bigl\{
I,X,Y,Z,\allowbreak
R_X\!\left(\frac{\pi}{2}\right),\allowbreak
R_Y\!\left(\frac{\pi}{2}\right),\allowbreak 
R_Z\!\left(\frac{\pi}{2}\right),\allowbreak 
H,\allowbreak 
R_Z\!\left(\frac{\pi}{2}\right)X,\allowbreak
R_X\!\left(\frac{\pi}{2}\right)Y
\bigr\}$,
where \(R_\alpha(\theta)\) denotes a rotation through angle \(\theta\) about the axis \(\alpha\). $X,Y$ and $Z$ denote the Pauli operators and $H$ is the Hadamard operation.
These are supplemented by three ancilla-assisted instruments resulting in six non-deterministic linearly independent operations. 
For each such instrument, the ancillary qubit is prepared in the positive eigenstate of one of the three nontrivial Pauli operators, \(X\), \(Y\), or \(Z\). Then, a SWAP operation is applied between the system and the ancilla, and the ancilla is subsequently measured in the corresponding Pauli basis. These deterministic and non-deterministic operations together form a set of linearly independent operations whose Choi representations span the local operation space required for informational completeness for the qubit intermediate lab $B$.
Note that the particular choice of operations is not unique. 
Any set whose Choi representations span the required local operator space serves the purpose. 

Following lab $B$, the system undergoes a second free evolution for time $d_2$,
after which measurements are performed at laboratory $C$ using an overcomplete set of observables $\{X, Y, Z\}$.
The two idle intervals \(d_1\) and \(d_2\) are implemented using the native timing resolution of the IBM processor, with \(1\,dt=4\,\mathrm{ns}\). 
Although no explicit control operation is applied during the delay intervals, the system remains coupled to uncontrolled degrees of freedom and continues to undergo open-system evolution.

The delay intervals are varied between $1$ and $95\,dt$ across the dataset. For a given system qubit and a fixed delay pair $(d_1,d_2)$, the complete combination of preparations, intermediate operations, and measurements gives rise to $156$ distinct circuits.
Each circuit is executed with 2000 shots, from which the 
outcome probabilities are obtained. These probabilities form the experimental data used subsequently to reconstruct the underlying process via maximum likelihood estimation and in our analysis using machine learning methods.

\subsection{Chip level experimental design}
Each tomography experiment is associated with a system qubit $q_S$ and an ancillary qubit $q_A$, the latter being used to implement nondeterministic operations. To facilitate direct two-qubit operations between the system and the ancilla, the two qubits are selected to be adjacent in the device connectivity graph. This connectivity information is obtained from the backend coupling map.

To make efficient use of limited hardware access time and the availability of multiple qubits on a superconducting processor, we implement tomography experiments in parallel across the device. However, simultaneous operations on neighboring qubits may introduce additional noise through crosstalk \cite{Gambetta2012,Ketterer2023,VallesSanclemente2026}. We therefore adopt a hardware-aware parallelization strategy aimed at minimizing unwanted interactions between concurrently executed tomography experiments.

To this end, we impose a one-hop isolation criterion based on the device connectivity graph. For each system--ancilla pair, an additional neighboring qubit of the system is reserved and kept idle. All qubits outside this allocation that are directly connected to any of the allocated qubits are also kept idle throughout the experiment. Additionally, any remaining unallocated qubits are left idle.

Qubits are selected using a greedy algorithm based on the device connectivity graph. Qubit allocations are constructed subject to the connectivity and one-hop isolation criteria described above. The allocated qubits and their immediate neighbors are excluded from subsequent allocations, although the respective one-hop neighborhoods may overlap. To accommodate a large number of parallel experiments, qubits are considered in ascending order of their degree in the connectivity graph, where the degree denotes the number of directly connected neighbors. This lowest-degree-first strategy prioritizes less connected qubits with the aim of preserving flexibility for subsequent allocations.

Using this approach, we identify $32$ system--ancilla pairs on the processor, enabling the data acquisition of $32$ multi-time process tomography experiments simultaneously in a run. Across $72$ such experimental runs, a total of $2304$ multi-time processes are characterized, forming the complete dataset used in this work.

\subsection{Reconstruction technique}
\subsubsection{MLE with IC data}

{ In order to obtain the estimate of non-Markovianity from the tomography data, we use maximum-likelihood estimation to reconstruct a physical process, detail of which is outlined in the following. For each experimental setting $\mathbf{x}=(x_1,x_2,x_3)$, consider $\rho_{x_1}$ represents the preparation at laboratory $A$, $\mathcal{M}_{b|x_2}$ an outcome map of the instrument applied at $B$, and $E_{c|x_3}$ a measurement effect at $C$. These experimentally implemented operations are treated as fixed during reconstruction. The joint outcome probabilities are for a given process $W$ is obtained through the generalised Born rule as $p_{bc|\mathbf{x}}(W)
=
\operatorname{Tr}\!\left[W \mathcal{T}_{bc|\mathbf{x}}\right],$
where the testor $\mathcal{T}_{bc|\mathbf{x}}$ is \
\begin{align}
\mathcal{T}_{bc|\mathbf{x}}
=
\rho_{x_1}^{T}
\otimes {M}_{b|x_2}^{T}
\otimes E_{c|x_3},
\end{align}
where $M_{b|x_2}$ is the Choi representation of $\mathcal{M}_{b|x_2}$ given by $
M_{b|x_2}
=
\sum_{j,k}|j\rangle\langle k|
\otimes
\mathcal{M}_{b|x_2}(|j\rangle\langle k|)
$.  Consider $n_{bc|\mathbf{x}}$ be the observed counts for the outcome $bc$ for setting $\mathbf{x}$ and $N_{\mathbf{x}}=\sum_{b,c}n_{bc|\mathbf{x}}$ the total number of shots for setting $\mathbf{x}$. Assuming independent circuit repetitions, the multinomial likelihood, up to factors independent of $W$, is
\begin{align}
\mathcal{L}(W)
\propto
\prod_{\mathbf{x},b,c}
p_{bc|\mathbf{x}}(W)^{n_{bc|\mathbf{x}}}.
\end{align}
MLE selects a valid process that maximises this likelihood. Since the logarithm is strictly increasing, maximising $\mathcal{L}(W)$ is equivalent to maximising its logarithm, or minimising the negative log-likelihood, i.e., 
\begin{align}
F(W)
=
-\sum_{\mathbf{x},b,c}
n_{bc|\mathbf{x}}\log p_{bc|\mathbf{x}}(W).
\end{align}
Using $f_{bc|\mathbf{x}}=n_{bc|\mathbf{x}}/N_{\mathbf{x}}$, this objective is equivalently the shot-weighted cross-entropy, namely $    F(W)
=
\sum_{\mathbf{x}}N_{\mathbf{x}}\,
H\!\left(f_{\mathbf{x}},p_{\mathbf{x}}(W)\right),$
where $H(f,p)=-\sum_{b,c}f_{bc}\log p_{bc}$ is the cross-entropy function. The reconstructed process is obtained through the following optimization} 
\begin{align*}
W_{\mathrm{MLE}}
\in\underset{W=W^\dagger}{\operatorname{arg\,min}}
\quad&
-\sum_{\mathbf{x},b,c}
n_{bc|\mathbf{x}}
\log\operatorname{Tr}\!\left[W \mathcal{T}_{bc|\mathbf{x}}\right]\\
\text{s.t.}\quad&
W\succeq0,
\\
&
\operatorname{Tr}_{C_1}W
=
R(W)\otimes I_{B_2},
\\
&
\operatorname{Tr}_{B_1}R(W)
=
I_{A_2},
\\
&
\operatorname{Tr}W
=
d_{A_2}d_{B_2}=4,
\end{align*}
where we have $R(W)=
\operatorname{Tr}_{B_2C_1}W/{d_{B_2}}.$
{The partial-trace constraints enforce causal ordering and normalization. The final trace constraint and is included to ensure normalization of probabilities. Importantly, the optimization is convex, i.e.,  the predicted probabilities depend linearly on $W$, the negative logarithm is convex, and the feasible set is the intersection of the positive-semidefinite cone with affine constraints. We use the MOSEK solver to perform the above MLE optimization. We note that for numerical stability, terms with zero observed count are omitted since a zero predicted probability for an outcome with positive count has infinite negative log-likelihood.}

It is important to note that MLE returns a unique estimate for a jointly informationally complete set of operations. Informationally incomplete data can admit multiple likelihood-maximizing processes, in which case MLE returns a physical process from the feasible set. 

\subsubsection{Restricted data and non-Markovianity bounds}
As mentioned above, data corresponding to informationally incomplete operations may correspond to several compatible processes. Here, we provide a procedure to search the family of compatible processes for minimum and maximum value of non-Markovianity. To this end, we normalize the process $w=W/4$, so that $\operatorname{Tr}w=1$. Defining the past and future blocks as $P=A_2B_1$ and $F=B_2C_1$, their reduced operators are
$w_P = \operatorname{Tr}_{F}w$, and
$w_F= \operatorname{Tr}_{P}w.$
The objective is given by
\begin{align}
\mathcal N(w)
&=
D\!\left(w\,\middle\|\,w_P\otimes w_F\right)
\\
&=
S(w_P)+S(w_F)-S(w),
\end{align}
where $S(\rho)=-\operatorname{Tr}(\rho\log \rho)$ is the von-Neumann entropy of the operator $\rho$. The product of these reduced operators is a physically valid Markovian process and is the closest one with respect to the relative-entropy distance \cite{Pollock2018b}.

For the processes considered in our work, the normalized feasible set is given by 
\begin{align}
\mathcal C_R
=
\Bigl\{
&w=w^\dagger:\;
w\succeq0,~ \operatorname{Tr}w=1, 
\nonumber\\
&
\operatorname{Tr}_{B_1}w_P
=
\frac{I_{A_2}}{2}, ~\operatorname{Tr}_{C_1}w
=
w_P\otimes\frac{I_{B_2}}{2},
\nonumber\\
&\qquad\qquad 4\operatorname{Tr}(w \mathcal{T}_r)
=
\widehat p_r
\quad\forall r\in R
\Bigr\}.
\end{align}
Here, $\widehat p_r$ are the restricted probabilities corresponding to the tester $\mathcal{T}_r$. 
predicted for the informationally complete maximum likelihood estimate process, rather than the empirical frequencies themselves. 
The extrema of interest are $\mathcal N_{\min}
=
\min_{w\in\mathcal F_R}\mathcal N(w),$  and $\mathcal N_{\max}
=
\max_{w\in\mathcal F_R}\mathcal N(w).$
Although $\mathcal{C}_R$ is a convex set, the objective function $\mathcal{N}$ is not a convex function. We therefore use successive entropy-based surrogate problems to obtain feasible candidates with decreasing or increasing non-Markovianity. For minimisation, we linearise the two concave marginal entropy terms around the current iterate $w^{(k)}$. For a positive-definite operator $x^{(k)}$, concavity gives
\begin{align}
S(x)
&\le
S\!\left(x^{(k)}\right)
+
\operatorname{Tr}\!\left[
G_x^{(k)}
\left(x-x^{(k)}\right)
\right],
\\
G_x^{(k)}
&=
-\log x^{(k)}-I.
\end{align}
Applying this inequality to both marginals yields an upper surrogate
\begin{align}
U_k(w)
={}&
S\!\left(w_P^{(k)}\right)
+
S\!\left(w_F^{(k)}\right)
-S(w)
\nonumber\\
&+
\operatorname{Tr}\!\left[
G_P^{(k)}
\left(w_P-w_P^{(k)}\right)
\right]
\nonumber\\
&+
\operatorname{Tr}\!\left[
G_F^{(k)}
\left(w_F-w_F^{(k)}\right)
\right].
\end{align}
This surrogate satisfies $\mathcal N(w)\le U_k(w)$, with equality at $w=w^{(k)}$. Omitting terms independent of the optimization variable, the next iterate is obtained from
\begin{align}
w^{(k+1)}
&=
\operatorname*{arg\,min}_{w\in\mathcal F_R}
\left\{
-S(w)+\operatorname{Tr}\!\left(G^{(k)}w\right)
\right\},
\\
G^{(k)}
&=
G_P^{(k)}\otimes I_F
+
I_P\otimes G_F^{(k)}.
\end{align}
This is a convex optimisation problem because negative entropy is convex and the remaining objective is linear. With an exact solution of the surrogate problem, the update obeys
\begin{align}
\mathcal N\!\left(w^{(k+1)}\right)
&\le
U_k\!\left(w^{(k+1)}\right)\\
&\le
U_k\!\left(w^{(k)}\right)
=
\mathcal N\!\left(w^{(k)}\right).\nonumber
\end{align}
Thus, this 
step produces monotone descent of the original objective.

For maximisation, we instead linearise the convex contribution $-S(w)$. Its tangent provides a lower bound,
\begin{align}
-S(w)
\ge
-S\!\left(w^{(k)}\right)
+
\operatorname{Tr}\!\left[
\left(\log w^{(k)}+I\right)
\left(w-w^{(k)}\right)
\right].
\end{align}
We additionally include a relative-entropy proximal term, giving the lower surrogate
\begin{align}
L_k(w)
={}&
S(w_P)+S(w_F)-S\!\left(w^{(k)}\right)
\nonumber\\
&+
\operatorname{Tr}\!\left[
\left(\log w^{(k)}+I\right)
\left(w-w^{(k)}\right)
\right]
\nonumber\\
&-
\tau D\!\left(w\,\middle\|\,w^{(k)}\right),
\qquad 
\end{align}
with $\tau>0.$
Since relative entropy is nonnegative, $L_k(w)\le\mathcal N(w)$, with equality at the current iterate. The proximal term discourages large changes and makes the surrogate strictly concave. After removing constant terms, the maximisation update becomes
\begin{align}
w^{(k+1)}
=
\operatorname*{arg\,max}_{w\in\mathcal F_R}
\Bigl\{
&S(w_P)+S(w_F)+\tau S(w)
\nonumber\\
&+
\operatorname{Tr}\!\left(H^{(k)}w\right)
\Bigr\},
\end{align}
where $H^{(k)}
=
(1+\tau)\log w^{(k)}+I.$
Maximising this concave surrogate over a convex feasible set is a convex optimisation problem. With an exact surrogate solution,
\begin{align}
\mathcal N\!\left(w^{(k+1)}\right)
&\ge
L_k\!\left(w^{(k+1)}\right)\\
&\ge
L_k\!\left(w^{(k)}\right)\nonumber
=
\mathcal N\!\left(w^{(k)}\right).
\end{align}
This step therefore produces monotone ascent.
For numerical implementation, the normalisation, causality and probability constraints are collected into an affine map $\mathcal A(w)=b$. Redundant constraints are removed using a singular-value decomposition in an isometric real coordinate representation of Hermitian matrices. The retained constraint operators are Hilbert--Schmidt orthonormal, which improves conditioning and simplifies affine corrections.

We solve the entropy subproblems through their smooth convex dual using L-BFGS-B. 
Moreover, to reduce dependence on initialisation, each search is performed from several feasible starting points. These include the maximum-entropy compatible process,
\begin{align}
w_{\mathrm{centre}}
=
\operatorname*{arg\,max}_{w\in\mathcal F_R}S(w),
\end{align}
a small mixture of the reference process with this centre, and additional states obtained by solving
\begin{align}
w_j^{(0)}
=
\operatorname*{arg\,max}_{w\in\mathcal F_R}
\left\{
\eta S(w)+\operatorname{Tr}(G_jw)
\right\},
\end{align}
where $\eta>0$, and $G_j$ are randomly generated Hermitian operators. These initialisations explore different parts of the compatible family while retaining its affine constraints.

The implementation assumes a numerically resolved full-rank feasible anchor so that the required matrix logarithms are well defined. Small eigenvalue floors are used for numerical evaluation of logarithms, and accepted candidates are checked for normalisation, positivity,The lowest and highest values among the returned feasible candidates are retained, with the reference process also included as a candidate. Consequently,
\begin{align}
\mathcal N_{\min}
\le
\mathcal N_{\mathrm{low}}^{\mathrm{found}}
\le
\mathcal N_{\mathrm{ref}}
\le
\mathcal N_{\mathrm{high}}^{\mathrm{found}}
\le
\mathcal N_{\max}.
\end{align}
Importantly,  multiple initialisations and convex surrogate solutions improve the numerical exploration but do not certify the global extrema of the original non-convex objective. The reported endpoints therefore quantify a range of compatible values found by the search. Their separation is a lower bound on the width of the full compatible range.

\subsection{Machine learning framework}
We formulate the estimation of non-Markovianity in the informationally incomplete regime as a supervised regression problem.

For each experimentally characterized process, the input vector $\mathbf{p}\in\mathbb{R}^{D}$ consists of the measured outcome probabilities retained by the chosen tester configuration, while the label is the corresponding reference value of $\mathcal{N}$ obtained from the informationally complete MLE reconstruction. To remove redundancy imposed by probability normalization, only $k-1$ outcome probabilities are retained as input features for each setting with $k$ possible outcomes. Thus, one and three probabilities are retained for binary- and four-outcome settings, respectively. The two tester configurations considered in the main text have input dimensions $D=36$ and $D=3$, respectively.

For the interpolation analysis, the complete dataset of 2304 processes is randomly partitioned into training, validation and test sets in a 60:20:20 ratio, with a different random partition generated for each seed. To assess generalization in the forecasting regime, we instead partition the data according to the total delay $d_{\mathrm{tot}}=d_1+d_2,$ such that processes with longer, previously unseen total delays are reserved for testing. These forecasting partitions are fixed across random seeds. Processes with $d_{\mathrm{tot}}\leq110 \ dt$ are used for training, 
those with $110 \ dt<d_{\mathrm{tot}}\leq140 \ dt$ form the validation set, 
and processes with $d_{\mathrm{tot}}>140 \ dt$ are used for testing. 
The delay values are used exclusively to construct these splits and are not supplied to the neural network as input features.
For each evaluation protocol and random seed, identical data partitions are used across all tester configurations.
Before training, each input feature is standardized using the mean and standard deviation computed from the corresponding training set, and the same transformation is subsequently applied to the validation and test sets. 
\\
\\
\textbf{Architecture:} A fully connected feed-forward artificial neural network, also known as a multilayer perceptron, is used to learn the mapping $\widehat{\mathcal{N}}=f_{\boldsymbol{\theta}}(\mathbf{p})$, where $\boldsymbol{\theta}$ denotes the trainable network parameters. For the $l$-th hidden layer, the transformation from its input to output activations is given by
\begin{equation}
    \mathbf{h}^{(l)}=\phi\left(\mathbf{W}^{(l)}\mathbf{h}^{(l-1)}+\mathbf{b}^{(l)}\right),
\end{equation}
where $\mathbf{W}^{(l)}$ and $\mathbf{b}^{(l)}$ are the trainable weight matrix and bias vector, respectively, $\mathbf{h}^{(l-1)}$ is the vector entering the layer $l$ and $\mathbf{h}^{(l)}$ is the vector produced by layer $l$. The initial vector $\mathbf{h}^{(0)}$ corresponds to the standardized input features. The network consists of four hidden layers with 256, 128, 64, and 32 neurons, respectively, each employing the rectified linear unit (ReLU) activation function, $\phi(z)=\max(0,z)$. The final hidden layer is connected to a single output neuron with a linear activation function, yielding the predicted non-Markovianity $\widehat{\mathcal{N}}$. The same hidden-layer architecture is used across all tester configurations and evaluation protocols, with the input dimension $D$ determined by the corresponding tester configuration.
\\
\\
\textbf{Training loss:} The trainable network parameters $\boldsymbol{\theta}$ are optimized by minimizing the mean squared error MSE between the predicted and reference non-Markovianity values. For a training set containing $N_{\mathrm{train}}$ processes, the loss function is defined as
\begin{equation}
    \mathcal{L}_{\mathrm{MSE}}(\boldsymbol{\theta})
    = \frac{1}{N_{\mathrm{train}}}
    \sum_{i=1}^{N_{\mathrm{train}}}
    \left(\mathcal{N}^{(i)}-\widehat{\mathcal{N}}^{(i)}\right)^2,
\end{equation}
where $\mathcal{N}^{(i)}$ and $\widehat{\mathcal{N}}^{(i)}$ denote the reference and predicted non-Markovianity values for the $i$-th process, respectively. The gradients of the loss with respect to $\boldsymbol{\theta}$ are computed using backpropagation. 
\\
\\
\textbf{Training procedure:} The network parameters are optimized using the adaptive moment estimation (Adam) algorithm with a learning rate of \(10^{-3}\). Training is performed for a maximum of \(500\) epochs using mini-batches of \(16\) samples, with the training data shuffled before each epoch. To limit overfitting, early stopping is employed by monitoring the validation MSE, with a patience of \(30\) epochs. The network parameters corresponding to the lowest validation MSE are restored for subsequent evaluation.

Training is repeated using four random seeds: $42$, $7$, $99$, and $123$. For the interpolation analysis, each seed determines both the data partition and the stochastic aspects of network training, including weight initialization and mini-batch shuffling, thereby assessing the robustness of the complete learning pipeline. In the forecasting analysis, the delay-defined data partitions remain fixed across seeds, such that the repetitions assess only stochastic variation in network training.
\\
\\

\textbf{Evaluation:} The predictive performance of the ANN is evaluated using the mean absolute error MAE and the coefficient of determination $R^2$. For a test set containing $N_{\mathrm{test}}$ processes, the MAE is defined as
\begin{equation}
\mathrm{MAE}
=
\frac{1}{N_{\mathrm{test}}}
\sum_{i=1}^{N_{\mathrm{test}}}
\left|
\mathcal{N}^{(i)}
-
\widehat{\mathcal{N}}^{(i)}
\right|,
\end{equation}
while the coefficient of determination is given by
\begin{equation}
R^2
=
1-
\frac{
\sum_{i=1}^{N_{\mathrm{test}}}
\left(
\mathcal{N}^{(i)}
-
\widehat{\mathcal{N}}^{(i)}
\right)^2
}{
\sum_{i=1}^{N_{\mathrm{test}}}
\left(
\mathcal{N}^{(i)}
-
\overline{\mathcal{N}}
\right)^2
},
\end{equation}
where $\overline{\mathcal{N}}$ denotes the mean reference value of non-Markovianity over the corresponding test set. The MAE quantifies the average magnitude of the prediction error, while $R^2$ measures the predictive performance relative to a constant predictor given by the mean of the reference values. Both metrics are computed separately on the corresponding test set for each run and reported as the mean $\pm$ standard deviation across the four runs.

\section*{Data availability}
The data supporting the findings of this study are available from the corresponding author upon reasonable request.
\section*{Code availability}
The code used to generate the results presented in this study is available from the corresponding author upon reasonable request.
\section*{Acknowledgments}

JK thanks Soumik Mahanti for many insightful discussions and acknowledges funding from Sydney Quantum Academy. The authors acknowledge the IBM Quantum Credits program for providing access to the quantum hardware used in this work. The views expressed in this work are those of the authors and do not necessarily reflect the official policy or position of IBM or the IBM Quantum team. 

\bibliography{sn-bibliography}

\end{document}